\documentclass[twocolumn,floatfix,prx,aps,showpacs]{revtex4-1}
\usepackage{graphicx,amsmath,amssymb,color}
\usepackage{nicefrac}
\usepackage[titletoc,title]{appendix}

\newcommand{\be}{\begin{equation}}
\newcommand{\ee}{\end{equation}}

\newcommand{\ba}{\begin{eqnarray}}
\newcommand{\ea}{\end{eqnarray}}

\renewcommand{\vec}[1]{{\textbf{\textit{#1}}}}

\begin{document}

\title{The nature of composite fermions and the role of particle hole symmetry: A microscopic account}
\author{Ajit C. Balram and J. K. Jain}
\affiliation{
Department of Physics, 104 Davey Lab, Pennsylvania State University, University Park, Pennsylvania 16802, USA} 
   
\begin{abstract} 
Motivated by the issue of particle-hole symmetry for the composite fermion Fermi sea at the half filled Landau level, Dam T. Son has made an intriguing proposal [Phys. Rev. X {\bf 5}, 031027 (2015)] that composite fermions are Dirac particles. We ask what features of the Dirac-composite fermion theory and its various consequences may be reconciled with the well established microscopic theory of the fractional quantum Hall effect and the 1/2 state, which is based on {\em non-relativistic} composite fermions. Starting from the microscopic theory, we derive the assertion of Son that the particle-hole transformation of electrons at filling factor $\nu=1/2$ corresponds to an effective time reversal transformation (i.e. $\{\vec{k}_j\}$$\rightarrow$$\{-\vec{k}_j\}$) for composite fermions, and discuss how this connects to the absence of $2k_{\rm F}$ backscattering in the presence of a particle-hole symmetric disorder. By considering bare holes in various composite-fermion $\Lambda$ levels (analogs of electronic Landau levels) we determine the $\Lambda$ level spacing and find it to be very nearly independent of the $\Lambda$ level index, consistent with a parabolic dispersion for the underlying composite fermions. Finally, we address the compatibility of the Chern-Simons theory with the lowest Landau level constraint, and find that the wave functions of the mean-field Chern-Simons theory, as well as a class of topologically similar wave functions, are surprisingly accurate when projected into the lowest Landau level.  These considerations lead us to introduce a ``normal form" for the unprojected wave functions of the $n/(2pn-1)$ states that correctly capture the topological properties even without lowest Landau level projection.
\end{abstract}
\maketitle

\section{Introduction}
Ever since the discovery of the phenomenon of the fractional quantum Hall effect (FQHE) \cite{Tsui82}, the system of electrons confined to two dimensions and exposed to a strong magnetic field has served a playground for the study of complex yet elegant new structures emerging as a result of interactions. The vast phenomenology of the lowest Landau level (LLL) has been securely explained or predicted in terms of the emergence of topological particles called composite fermions, which are bound states of electrons and quantized vortices \cite{Jain89,Lopez91,Halperin93,Jain07}. Composite fermions experience a reduced effective magnetic field and can be treated as non-interacting to a first approximation. Prominent among the successes of the composite (CF) theory are the FQHE at the Jain fractions $\nu=n/(2pn\pm 1)$ ($p$ is a positive integer), which are explained as $\nu^*=n$ integer quantum Hall effect (IQHE) of composite fermions; the compressible state at $\nu=1/2$, which is understood as the Halperin-Lee-Read (HLR)\cite{Halperin93} Fermi sea of composite fermions in vanishing effective magnetic field (also see Kalmeyer and Zhang \cite{Kalmeyer92}); the spin polarization of the FQHE states \cite{Du95,Park98,Kukushkin99,Yeh99,Bishop07,Padmanabhan09,Feldman13,Liu14,Balram15a}; the FQHE at 5/2\cite{Willett87}, believed to be described by the Moore-Read Pfaffian wave function of the chiral p-wave paired state of composite fermions\cite{Moore91,Read00}; and various charged and neutral excitations that are understood as excitations of composite fermions across their Landau-like levels called $\Lambda$ levels ($\Lambda$Ls) \cite{Dev92,Jain98,Scarola00,Kukushkin00,Mandal01,Kang01,Dujovne03,Dujovne05,Gallais06,Kukushkin07,Kukushkin09,Majumder09,Rhone11,Wurstbauer11,Balram13}.  

More recently, FQHE has also been observed for Dirac electrons in graphene\cite{Xu09,Bolotin09,Feldman13,Amet15}.  Interestingly, to the extent the finite width and Landau level (LL) mixing effects are negligible, the FQHE physics in the $n=0$ LL of non-relativistic electrons in GaAs and in the $n=0$ LL of Dirac electrons in graphene are identical \cite{Apalkov06,Nomura06,Toke06}. The nature of the emergent composite fermions thus does not depend on whether the parent electrons are relativistic Dirac electrons or non-relativistic electrons obeying a parabolic dispersion. (The physics in the $n\neq 0$ LLs of the Dirac and non-relativistic systems are different\cite{Apalkov06,Nomura06,Toke06,Shibata09,Balram15c}, but that is not relevant to the issue at hand.)

Tremendous excitement has recently been generated by an ingenious proposal of Son \cite{Son15} wherein he models composite fermions as Dirac particles, akin to those on the surface of a three-dimensional (3D) topological insulator. His motivation for introducing ``Dirac composite fermions" comes from the following observation. A widely employed approach for incorporating the CF physics into a theoretical framework is through the Chern-Simons (CS) field theory, developed by Lopez and Fradkin\cite{Lopez91} and HLR\cite{Halperin93}, in which one performs a singular gauge transformation to map the electron system into that of composite fermions at an effective magnetic field.  One then calculates (in an approximate scheme) quantities for composite fermions at the effective magnetic field, and then uses a dictionary for translating these quantities into those for electrons, which are what the experiments measure. One of the problems of the CS approach is that it does not impose the LLL constraint, and thus does not satisfy particle-hole (p-h) symmetry at half filling, which is an exact symmetry of the Hamiltonian describing electrons interacting via a two-body interaction while being confined to the LLL. 

Son begins by noting that the physics of the half filled $n=0$ LL in GaAs is identical to that of the half filled $n=0$ LL in a Dirac system. P-h symmetry arises differently in the two models. In GaAs, p-h symmetry emerges at $\nu=1/2$ only in the limit when admixture with higher LLs is suppressed, whereas in the Dirac system, p-h symmetry at $\nu=1/2$ is valid even in the presence of LL mixing, because the LLs are symmetrically located above and below the Dirac point. Going to a system of Dirac electrons thus decouples LLL projection and p-h symmetry. Son then proposes that a p-h symmetric description of the CF Fermi sea is obtained effectively in terms of a Fermi sea of electrically neutral Dirac composite fermions coupled to an emergent dynamical gauge field. He conjectures that the p-h symmetry of electrons maps into time reversal symmetry of the Dirac composite fermions. One of the principal properties of the Dirac sea, as opposed to the HLR Fermi sea, is that an adiabatic loop around the Fermi circle produces a Berry phase of $\pi$.  (A $\pi$ Berry phase for the $\nu=1/2$ CF Fermi sea can also be motivated from the work of Haldane \cite{Haldane04} who demonstrated a  connection between the intrinsic anomalous Hall effect and the topological Berry phase around the Fermi circle.)

A hallmark of the CF theory is the deep connection it reveals between the compressible state at $\nu=1/2$ and the incompressible FQHE states at $\nu=n/(2n\pm 1)$, demonstrating that they all have the same physical origin. The theory of the 1/2 Fermi sea must also explain the fractions $\nu=n/(2n\pm 1)$, and {\em vice versa.} Son of course appreciates this and has addressed the issue of FQHE within his Dirac CF approach. In his approach, when the filling factor is varied away from half filling, the Dirac composite fermions experience the standard effective magnetic field, and exhibit FQHE at the Jain fractions. A radical step of the Dirac-CF viewpoint is to abandon the definition of the composite fermion as the bound state of an electron and an even number of quantized vortices\cite{Levin16}. This is necessitated because the number of Dirac composite fermions is, in general, not equal to the number electrons but to half the number of flux quanta penetrating the sample. In particular, the fraction $\nu=n/(2n+1)$ and its hole counterpart at $\nu=(n+1)/(2n+1)$ map into $\nu^*=n+1/2$ ``IQHE" of Dirac composite fermions in positive and negative effective magnetic fields, respectively. (Here, ``IQHE" refers to QHE of non-interacting fermions filling an integer number of Landau levels, even though the Hall resistance for Dirac fermions is half-integrally quantized.) Thus, even though the issue of p-h symmetry of the HLR Fermi sea served as the initial motivation of the Dirac-CF view, it entails implications for the physical meaning of composite fermions as well as the general structure of the CF theory. 

Son's proposal has stimulated much further work\cite{Wang15b,Metlitski15,Wang15,Geraedts15,Potter15,Murthy15,Liu15a,Metlitski15a,Mross15,Kachru15,Mulligan16,Levin16}.  Aiming to clarify its conceptual underpinnings, Wang and Senthil \cite{Wang15b,Wang15} and Metlitski and Vishwanath \cite{Metlitski15} have argued, with further justification by Mross, Alicea and Motrunich \cite{Mross15}, that the system of Dirac electrons at the surface of a 3D topological insulator coupled to the electromagnetic field is dual to a system of neutral fermions coupled to an emergent gauge field whose flux is proportional to the electron density, in the sense that the two theories live in the same Hilbert space. In light of this duality, the authors argue that it is natural to identify the Fermi sea state at the half filled $n=0$ LL of the former with the Dirac Fermi sea of the latter at zero magnetic field. Wang and Senthil also provide a picture for how two-component nature of composite fermions may emerge as a result of LLL projection \cite{Wang15}. Murthy and Shankar have extended their Hamiltonian theory \cite{Murthy03} to Dirac composite fermions\cite{Murthy15}. Geraedts {\em et al.}\cite{Geraedts15} provide a convincing demonstration, in a DMRG calculation, of an absence of $2k_{\rm F}$ backscattering for the CF Fermi sea at $\nu=1/2$ in the presence of a particle-hole symmetric disorder. This has been taken as an evidence for the Dirac nature of composite fermions. Interestingly, the issue of transport at $\nu=1/2$ in a random flux disorder (which simulates density variations of composite fermions) had already been investigated by Kalmeyer and Zhang \cite{Kalmeyer92} and by Kalmeyer {\em et al.} \cite{Kalmeyer93} in the early 1990s, who find an absence of localization for disorder that preserves time reversal symmetry. 

The view that composite fermions are Dirac fermions presents a paradox, however. To see this, we recall that there also exists a microscopic theory of non-relativistic composite fermions \cite{Jain89,Jain07}, which proceeds by constructing explicit wave functions for the ground and excited states at arbitrary fillings by analogy to weakly interacting {\em non-relativistic} fermions in an effective magnetic field. The validity of the microscopic theory is beyond dispute. In particular, Son's criticisms of the Chern-Simons field theoretical formulation that led him to introduce Dirac composite fermions are not applicable to the microscopic theory. This implies that the deficiencies of the CS theory are not necessarily an evidence against the non-relativistic nature of composite fermions per se. From the perspective of the microscopic theory, there is no reason to question the non-relativistic nature of composite fermions. 

Even though an effective theory aims to capture the long-distance low-energy physics, one expects it to be compatible with, and hopefully even derivable from, the microscopic theory. This raises a number of questions. What assertions of the Dirac-CF picture are consistent with the microscopic theory? Can one derive, microscopically, the absence of $2k_{\rm F}$ backscattering for composite fermions at $\nu=1/2$ for a p-h symmetric disorder? How is the $\pi$ Berry phase for the Fermi circle related to these issues, and how, if at all, does it follow from the microscopic theory? What is the dispersion of  the composite fermion? Are the deficiencies of the CS theory intrinsic, or are they of a technical nature? Is the CS theory compatible with LLL projection? If so, does it produce microscopically accurate results and a p-h symmetric CF Fermi sea at $\nu=1/2$?  We address all of these issues in this article.  Our philosophy below will be always to frame the discussion entirely in terms of the wave functions of electrons confined to the LLL, and then ask what the results mean in terms of composite fermions. We again stress that the nature of the parent electrons, whether they are non-relativistic or Dirac, plays no role in our considerations.

The plan of the article is as follows. In Sec. \ref{sec2}, we summarize the microscopic theory of non-relativistic composite fermions and discuss how several criticisms of the CS theory do not apply to the microscopic theory. In Sec. \ref{sec3} we provide a derivation, using the microscopic theory, of Son's insight that at $\nu=1/2$ p-h transformation on electrons acts as a time reversal-like transformation $\{\vec{k}_j\}$$\rightarrow$$\{-\vec{k}_j\}$ of composite fermions. We show that this implies absence of $2k_{\rm F}$ backscattering of a single composite fermion in the presence of a p-h symmetric perturbation. We also discuss the relation of this result to a $\pi$ Berry phase for the CF Fermi sea. In Sec.~\ref{sec4} we ask if the excitation spectrum can distinguish between the non-relativistic and Dirac CF descriptions. We  calculate the spacing between the emergent $\Lambda$Ls of composite fermions by considering bare (undressed) CF holes in various $\Lambda$ levels, and find that the $\Lambda$Ls are equally spaced, as expected for composite fermions with a parabolic dispersion. We investigate in Sec.~\ref{sec5} if the CS mean-field formulation of composite fermions is compatible with LLL projection and, if so, whether it provides a p-h symmetric description at $\nu=1/2$. We define LLL projection in an operational sense within the CS approach, and find that it leads to a reasonably good p-h symmetric description. In Sec.~\ref{LLLgen} we note that the CS mean field wave functions are part of a more general class of topologically similar wave functions, which all produce reasonably good wave functions when projected into the LLL. However, as noted in Sec.~\ref{sec6}, these are not all sensible wave functions without LLL projection.  We consider the issue of the topological local charge of the quasiparticle excitations in the Chern-Simons mean field states, and introduce a ``normal" form for the unprojected composite fermions that has many nice features. The article is concluded in Sec.~\ref{sec7}.

\section{Microscopic theory}
\label{sec2}

The fundamental principle of the CF theory\cite{Jain89,Lopez91,Halperin93,Jain07} is that interacting electrons at a filling factor $\nu=\nu^*/(2p\nu^*\pm 1)$ transform into weakly interacting (non-relativistic) composite fermions at $\nu^*$. One consequence of this is that the low energy spectrum of the former, which results entirely from interelectron interactions, resembles the known low energy spectrum of non-interacting fermions at $\nu^*$. In particular, at $\nu^*=n$, the latter produces an unique incompressible IQHE state, which implies FQHE at $\nu=n/(2pn\pm 1)$. The CF theory goes beyond a qualitative explanation of the phenomenology by constructing explicit wave functions\cite{Jain89,Jain07} for interacting electrons in the LLL in terms of the known wave functions of non-relativistic fermions $\Phi_{\nu^*}^{\gamma}$:
\begin{equation}
\Psi_{\nu=\frac{\nu^*}{2p\nu^*\pm 1}}^{{\rm CF}, \gamma}={\cal P}_{\rm LLL}\prod_{j<k}(z_j-z_k)^{2p} \Phi_{\pm \nu^*}^{\gamma}
\label{Jainwf}
\end{equation}
Here $\gamma$ labels the different eigenstates, $\Phi_{-\nu^*}^{\gamma}=[\Phi_{\nu^*}^{\gamma}]^*$, ${\cal P}_{\rm LLL}$ denotes the LLL projection, and $z_j=x_j-iy_j$ denotes the position of the $j$th electron as a complex number. This single equation gives wave functions, and thus energies, for all low energy eigenstates at arbitrary filling $\nu$ in the LLL, and thereby subjects itself to rigorous and non-trivial qualitative and quantitative tests.  In this wave function, the factor $\Phi_{\pm \nu^*}^{\gamma}$ contains the Gaussian factor $\exp[-\sum_j|z_j|^2/4\ell^2]$ where $\ell=\sqrt{\hbar c/eB}$ is the magnetic length at the {\em external} magnetic field $B$. 
In the limit of $n\rightarrow \infty$, for $p=1$, the above wave functions reduce to the CF Fermi liquid state at $\nu=1/2$. Here the wave function is given by\cite{Rezayi94}
\begin{equation}
\Psi_{\nu=1/2}^{\rm CF}={\cal P}_{\rm LLL}\prod_{j<k}(z_j-z_k)^{2} \Phi_{\rm FS}(B^*=0) e^{-\sum_j|z_j|^2/4\ell^2}
\label{CFFS}
\end{equation}
where $\Phi_{\rm FS}(B^*=0)$ is the wave function of the Fermi sea of non-relativistic fermions and the Gaussian factor is displayed explicitly.

In the microscopic theory, even though one uses inspiration from the CF physics, the final wave functions are written for electrons, which allows calculations of various quantities directly for electrons in the actual magnetic field.  The above wave functions have been tested in painstaking detail in the entire filling factor range in the LLL where FQHE is seen and shown to be essentially exact representations of the Coulomb solutions for the ground as well as excited states \cite{Dev92,Rezayi94,Jain97,Jain07,Balram13}. These studies establish a direct relation between the incompressible FQHE states at $\nu=n/(2pn\pm 1)$ and the compressible state at $\nu=1/2$ with IQHE and Fermi sea of non-relativistic fermions, respectively.

It is worth noting that several deficiencies of the CS field theory, including the ones that served as a motivation for the introduction of Dirac composite fermions, do not apply to the microscopic theory of non-relativistic composite fermions. 

(i) {\em Energy scale}: In the CS theory the energy scale at the mean field level is set by the cyclotron energy, and not by the Coulomb energy as expected for a theory confined to the LLL. On the other hand, Jain's wave functions of Eq.~\ref{Jainwf} reside, by construction, in the LLL, and thus produce various energies in the Coulomb units. 

(ii) {\em Nature of composite fermions}: In the CS theory, composite fermions are modeled as bound states of electrons and point flux quanta. These are actually strongly interacting due to the gauge interaction between them, and not the final quasiparticles. 

In the microscopic theory, composite fermions are viewed as the bound states of electrons and an even number of quantized vortices. These are not identical to the composite fermions of the CS theory, but the two are topologically similar, in the sense that they produce the same winding phases. The binding of quantized vortices to electrons is most evident in the ``unprojected" wave functions $\Psi_{\nu=\frac{n}{2pn\pm 1}}^{\rm CF-un}=\prod_{j<k}(z_j-z_k)^{2p} \Phi_{\pm n}$. The act of projecting these wave functions into the LLL turns the composite fermions into more complicated objects, but it can be argued that the LLL projected composite fermions are adiabatically connected to the unprojected composite fermions, thus providing an adiabatic justification for the definition of composite fermions as bound states of electrons and vortices even within the LLL. (This is further discussed in Sec.~\ref{sec6}.) A strong evidence for an adiabatic connectivity between the unprojected and the projected composite fermions is that, as seen in exact spectra evaluated in the LLL, the lowest band at $\nu$ has a one-to-one correspondence with that at $\nu^*$, suggesting that the low energy structure does not change upon LLL projection. Furthermore, for the 2/5 state, it is possible to construct an adiabatic scheme \cite{Rezayi91} that interpolates between the unprojected and the projected Jain wave functions, starting from a model that produces the unprojected wave function as the exact ground state\cite{Jain90}, confirming that the gap does not close during the adiabatic process \cite{Rezayi91}. We note that the almost perfect agreement with the exact Coulomb eigenstates demonstrates that the LLL projected composite fermions are essentially the true composite fermions that occur in nature.

(iii) {\em Electron-based composite fermions vs. hole-based composite fermions}: In the CS theory the flux quanta are attached to electrons. It is not possible to attach flux quanta to the holes of the LLL. The CS theory thus explicitly breaks p-h symmetry, as also indicated by the absence of LLL projection in this theory.  Within the microscopic theory, however, vortices may be attached to either electrons or holes. When vortices are attached to holes, the unprojected wave function is not meaningful, but its LLL projection yields a physical wave function. {\em A priori}, the relation between the two constructions, with vortices attached to electrons or holes, is unclear. Consider, for example, the CF Fermi sea wave function of Eq.~\ref{CFFS}. One can construct two Fermi sea wave functions here, one obtained by attaching vortices to electrons and the other to holes. Are they two distinct CF Fermi seas? Explicit calculations by Rezayi and Haldane\cite{Rezayi00} in the torus geometry have demonstrated that the two wave functions are essentially identical. More specifically, for $N=16$ particles, the wave function constructed above has an overlap of 0.9994 with its hole partner (which is essentially the Fermi sea wave function constructed from the holes in the LLL). This shows that the two descriptions are equivalent. 

In an analogous fashion, at each fraction of the form $n/(2n\pm 1)$ wave functions can be constructed using either electron-based composite fermions or hole-based composite fermions. As discussed in more detail in Section \ref{sec3}, there is good evidence that these actually represent the same state.

(iv) {\em Particle-hole symmetry of the CF Fermi sea}: The CF Fermi sea in the HLR theory is not p-h symmetric, as it is not constrained to the LLL. However, the microscopic wave function of the $\nu=1/2$ CF Fermi sea has been found to satisfy the p-h symmetry to a high accuracy in computer calculations \cite{Rezayi00}, as mentioned above.  The DMRG calculations of Zaletel {\em et al.}\cite{Zaletel15} also suggest that the p-h symmetry is not spontaneously broken at $\nu=1/2$. 

We stress, parenthetically, that it is impossible to rule out a spontaneous braking of particle-hole symmetry at $\nu=1/2$. One way this could happen would be if, eventually, the CF Fermi sea becomes unstable to a pairing that is too subtle to be captured by the finite systems being studied. Should such pairing be of the Moore-Read kind \cite{Moore91}, then that would suggest a spontaneous breaking of p-h symmetry\cite{Levin07,Lee07,Wojs10,Rezayi11,Peterson14,Zaletel15,Pakrouski15,Tyler15}. In this case, one can imagine two distinct CF Fermi seas which are the normal states of the Pfaffian and the anti-Pfaffian paired CF states.  We refer the reader to the article by Barkeshli, Mulligan and Fisher~\cite{Barkeshli15} for the experimental consequences of a spontaneously broken particle-hole symmetry at $\nu=1/2$. In what follows, we will assume that the state at $\nu=1/2$ is particle-hole symmetric.  We note that the Dirac-CF postulates also assumes a unique p-h symmetric CF Fermi sea. 

(v) {\em The paradox of CF Hall conductivity}:  Kivelson {\em et al.} \cite{Kivelson97} considered the 1/2 state and noted that in the presence of a nonzero p-h symmetric disorder ($\sigma_{xx} \neq 0$), the correct electronic Hall conductivity is $\sigma_{xy}=e^2/2h$, which requires that the CF Hall conductivity must be equal to $\sigma_{xy}^{\rm CF}=-(1/2) e^2/h$, according to the standard formulae of the CS theory relating the CF and the electronic quantities. However, at the mean field level, composite fermions do not experience any effective magnetic field at $\nu=1/2$, which produces $\sigma_{xy}^{\rm CF}=0$, and it is not known what corrections to the mean field theory would produce $\sigma_{xy}^{\rm CF}=-(1/2) e^2/h$ within the CS approach. This paradox has to do with translating between the CF and electron transport coefficients; this paradox is absent in the the microscopic formulation that deals directly with electronic states (although an actual calculation of the corrections to the ideal transport coefficients due to the presence of disorder and LL mixing would be highly nontrivial within the microscopic theory).

(vi) {\em CF Fermi wave vector}:
Within the mean field CS approach, it would appear that the Fermi wave vector of the CF Fermi sea slightly away from $\nu=1/2$ would nominally be given by $k_{\rm F}^*=\sqrt{4\pi \rho_{\rm e}}$, where $\rho_{\rm e}$ is the electron density. Beautiful experiments from Kamburov {\em et al.}\cite{Kamburov14b} have found that the CF Fermi wave vector is given by $k_{\rm F}^*=\sqrt{4\pi \rho_{\rm e}}$ for $\nu<1/2$ but by $k_{\rm F}^*=\sqrt{4\pi \rho_{\rm h}}$ for $\nu>1/2$, where $\rho_{\rm h}$ is the density of holes in the LLL.  Ref.~\cite{Balram15b} showed that no such paradox appears if the CF Fermi wave vector is evaluated using the {\em microscopic} theory. First of all, it can be shown that for two states at $\nu$ and $1-\nu$ related by p-h symmetry, $k_{\rm F}^*\ell$ is the same (where $k_{\rm F}^*$ is defined through the Friedel oscillations in the pair correlation function). Second, because Jain's wave functions at $n/(2n+1)$ and $1-n/(2n+1)=(n+1)/(2n+1)$ are related by p-h symmetry to an excellent approximation, they produce the same $k_{\rm F}^*\ell$.  An explicit calculation \cite{Balram15b} shows that the Fermi wave vector is close to the nominal value $k_{\rm F}^*=\sqrt{4\pi \rho}$ with the density $\rho$ taken equal to that of the minority carriers, consistent with the experimental observation \cite{Kamburov14b}.

Motivated by the Princeton experiments\cite{Kamburov14b}, Barkeshli, Mulligan and Fisher~\cite{Barkeshli15} have put forth the postulate that the electron-based CF Fermi sea and the hole-based CF Fermi sea are distinct phases of matter at $\nu=1/2$, and a spontaneous breakdown of p-h symmetry selects one of them. 
Building on this work Mulligan, Raghu and Fisher~\cite{Mulligan16} show how p-h symmetry can emerge in the half-filled Landau level by considering a system of alternating quasi-one-dimensional strips of electron-based CF Fermi sea and hole-based CF Fermi sea.

\section{Particle-hole symmetry and the $1/2$ CF Fermi sea}
\label{sec3}

As noted above, one of the predictions of the Dirac CF description is that a closed loop around the CF Fermi sea produces a $\pi$ Berry phase, which has measurable experimental consequences \cite{Son15,Potter15}.  Numerical DMRG studies\cite{Geraedts15} have nicely confirmed absence of backscattering at $\nu=1/2$ in the presence of a particle-hole symmetric disorder. In view of the earlier numerical results \cite{Rezayi00}, there is hardly any doubt that the DMRG studies are generating, in the absence of disorder, a 1/2 wave function that is very close to the wave function of Eq.~\ref{CFFS}, and thus exploring its properties. 

Our approach will be to frame the discussion entirely in terms of the electron wave functions, and interpret the results in terms of composite fermions. We first identify what p-h transformation for electrons means in terms of composite fermions. We show that at $\nu=1/2$ it corresponds to the time reversal transformation $\{\vec{k}_j\}$$\rightarrow$$\{-\vec{k}_j\}$ of composite fermions, consistent with Son\cite{Son15}. We then discuss what it implies for backscattering of composite fermions and also for the Berry phase for an adiabatic loop around the Fermi circle.

\subsection{Particle hole transformation in CF theory}

We first consider the issue of p-h transformation in the LLL. Let us consider $N$ electrons in $M$ LLL orbitals.  (We have in mind a compact geometry, such as the spherical geometry, so the LLL has a finite number of orbitals.)   A wave function $\Psi_\nu$ of electrons at filling factor $\nu$, in second quantized form, is given by
\begin{widetext}
\be
|\Psi_\nu\rangle = \int d^2 \vec{r}_1\cdots d^2 \vec{r}_N \Psi_{\nu}(\vec{r}_1,\cdots, \vec{r}_N) \hat{\psi}^\dagger(\vec{r}_1)\cdots \hat{\psi}^\dagger(\vec{r}_N)|0\rangle
\label{fockwf}
\ee
where $\Psi_{\nu}(\vec{r}_1,\cdots, \vec{r}_N)$ is the real space wave function,  $\hat{\psi}^\dagger(\vec{r})$ and $\hat{\psi}(\vec{r})$ are the standard LLL-projected creation and annihilation field operators\cite{Stone92}, and the state $|0\rangle$ refers to the empty state with no electrons.  We use either $z_j=x_j-i y_j$ or $\vec{r}_j=(x_j, y_j)$ to denote the position of the $j$th particle.

The p-h transformation, denoted by $\Theta$, is defined as:
\begin{eqnarray}
\Theta\,\Psi(\vec{r}_1,\cdots, \vec{r}_N) \Theta^{-1} &=& [\Psi(\vec{r}_1,\cdots, \vec{r}_N)]^*\\
\Theta\,\hat{\psi}^\dagger(\vec{r}) \Theta^{-1} &=& \hat{\psi}(\vec{r})\\
\Theta\,\hat{\psi}(\vec{r}) \Theta^{-1} &=& \hat{\psi}^\dagger(\vec{r})\\
\Theta\,|0\rangle &=& |\Phi_1\rangle
\end{eqnarray}
where $|\Phi_1\rangle$ denotes the state where the LLL is fully occupied:
\be
|\Phi_1\rangle=\int d^2 \vec{r}_1\cdots d^2 \vec{r}_M \Phi_1(\vec{r}_1,\cdots, \vec{r}_M) \hat{\psi}^\dagger(\vec{r}_1)\cdots \hat{\psi}^\dagger(\vec{r}_M)|0\rangle
\ee

With standard manipulations, it is straightforward to show that \cite{Girvin84,Girvin84b}
\be
\Theta\,|\Psi_\nu\rangle = \int d^2 \vec{r}_{N+1}\cdots d^2 \vec{r}_M \widetilde{\Psi}_{1-\nu}(\vec{r}_{N+1},\cdots, \vec{r}_M) \hat{\psi}^\dagger(\vec{r}_{N+1})\cdots \hat{\psi}^\dagger(\vec{r}_M)|0\rangle
\ee
where
\begin{equation}
\widetilde{\Psi}_{1-\nu}(\vec{r}_{N+1},\cdots, \vec{r}_M) = \int d^2 \vec{r}_1\cdots d^2 \vec{r}_N~ \Phi_1(\vec{r}_1,\cdots \vec{r}_M) 
\left[ \Psi_{\nu}(\vec{r}_{1},\cdots,\vec{r}_N)\right]^*
\label{Eqph}
\end{equation}
is the real space wave function for the p-h conjugate of $\Psi_{\nu}$. The p-h transformation is thus a complicated nonlocal transformation, requiring evaluation of multidimensional integrals.

We now show that for the FQHE states and their low energy excitations, the CF theory provides a relatively simple way of accomplishing the p-h transformation that does not require evaluating any multidimensional integrals. For this purpose, let us begin by considering the fully spin polarized state at $n/(2n+ 1)$.   Its wave function is given by 
\be
\Psi^{\rm CF}_{\frac{n}{2n+ 1}}(z_1,\cdots, z_N) = {\cal P}_{\rm LLL} \prod_{j<k=1}^N(z_j-z_k)^{2} \Phi_{n}(z_1,\cdots, z_N)\label{EqA}
\ee
Its p-h conjugate state $\widetilde{\Psi}_{\frac{n+1}{2n+ 1}}$ at $\nu=1-n/(2n+1)=(n+1)/(2n+1)$ can, in principle, be evaluated according to Eq.~\ref{Eqph}. However, the CF theory provides another way of constructing a state at that filling factor directly, by considering composite fermions at filling factor $\nu^*=n+1$ in negative effective magnetic field: 
\be
\Psi^{\rm CF}_{\frac{n+1}{2n+ 1}} (z_{N+1},\cdots, z_M) = 
{\cal P}_{\rm LLL}\prod_{j<k=N+1}^M(z_j-z_k)^{2} \Phi^*_{n+1}(z_{N+1},\cdots,z_M)
\label{EqB}
\ee
At first sight, the p-h conjugate $\widetilde{\Psi}_{\frac{n+1}{2n+ 1}}(z_{N+1},\cdots, z_M)$ obtained according to Eq.~\ref{Eqph} and $\Psi_{\frac{n+1}{2n+ 1}} (z_{N+1},\cdots, z_M)$ in Eq.~\ref{EqB} appear to be distinct, because in the former the vortices are attached to electrons whereas in the latter the vortices are attached to holes, suggesting very different correlations. From that vantage point, it is perhaps surprising, but nonetheless a fact, that these two wave functions do not represent distinct states but are dual descriptions of the same state, i.e., 
\be
\widetilde{\Psi}_{\frac{n+1}{2n+ 1}}(\vec{r}_{N+1},\cdots \vec{r}_{M}) \equiv
 \int d^2 \vec{r}_1\cdots d^2 \vec{r}_N~ \Phi_1(\vec{r}_1,\cdots \vec{r}_M) 
\left[ \Psi_{n\over 2n+1}(\vec{r}_{1},\cdots,\vec{r}_N)\right]^*
 \approxeq \Psi^{\rm CF}_{\frac{n+1}{2n+ 1}} (z_{N+1},\cdots, z_M) 
\label{EqAB}
\ee
\end{widetext}
This relation is supported by the following observations: (i) The two descriptions produce identical quantum numbers for the ground state as well as low energy excitations. (ii) The wave functions themselves are essentially identical. This has been tested explicitly for small systems, where both the wave functions can be constructed exactly on the computer\cite{Wu93,Davenport12}. For larger systems, we have computed the pair correlation function $g(r)$ for the wave functions at $\nu=(n+1)/(2n+1)$ and $\nu=n/(2n+1)$ and found that they accurately satisfy the relation expected for two states that are exactly related by p-h symmetry. See the supplementary section of Ref.~\cite{Balram15b} for further details. Eq.~\ref{EqAB} encapsulates how the CF theory provides a much simpler implementation of p-h transformation, without doing any integrals as in Eq.~\ref{Eqph}.  We stress, however, that the relation in Eq.~\ref{EqAB} is nontrivial, not exact, and does not apply to an arbitrary wave function but only to the low energy states of composite fermions. 

One of the corollaries of Eq.~\ref{EqAB} is that there is a unique incompressible state at each of the fractions $\nu=n/(2n\pm 1)$ for fully spin polarized electrons. This is consistent with the fact that only one state is found, both in experiments and in exact diagonalization studies, at each of these fractions (when the spin degree of freedom is frozen). The uniqueness of these states is also closely related to the uniqueness of the CF Fermi sea in the limit $n\rightarrow \infty$; otherwise the  states at $n/(2n-1)$ for composite fermions in negative effective magnetic field and those at $n/(2n+1)$ in positive effective magnetic field would produce different CF Fermi seas in the $n\rightarrow \infty$ limit.

The p-h transformation on electrons in the LLL should not be confused with p-h transformation of composite fermions. As a matter of fact, it is not possible to define p-h transformation for composite fermions because their Hilbert space is not restricted to their lowest $\Lambda$L. Nonetheless, Eqs. \ref{EqA}, \ref{EqB} and \ref{EqAB} tell us what transformation for composite fermions corresponds to p-h transformation of electrons. For now, we will replace the ``$\approxeq$" sign in Eq.~\ref{EqAB} by the ``$=$" sign, and come back to this assumption later. In a hopefully transparent short-hand notation, the p-h transformation of electrons corresponds, according to Eq.~\ref{EqAB}, to the transformation of composite fermions  as 
\be
\Theta\,\Phi_n(\vec{r}_1,\cdots, \vec{r}_N) \,\Theta^{-1}= [\Phi_{n+1}(\vec{r}_{N+1},\cdots, \vec{r}_M)]^*
\label{phtrans}
\ee
i.e., the state of $n$ filled $\Lambda$Ls of $N$ composite fermions in a positive effective magnetic field maps into the state of $n+1$ filled $\Lambda$Ls of $M-N$ composite fermions in a negative effective magnetic field (as indicated by complex conjugation).

One can similarly determine how the excitations transform under p-h transformation.  The particle excitation is an isolated composite fermion in an otherwise empty $\Lambda$L and the hole excitation is a missing composite fermion from an otherwise full $\Lambda$L. Straightforward application of the CF theory shows\cite{Jain07}:
\be
\Theta\,\Phi_n^{\rm hole}(\vec{r}_1,\cdots, \vec{r}_N) \,\Theta^{-1}= [\Phi_{n+1}^{\rm hole}(\vec{r}_{N+1},\cdots, \vec{r}_M)]^*
\ee
\be
\Theta\,\Phi_n^{\rm particle}(\vec{r}_1,\cdots, \vec{r}_N)\,\Theta^{-1}= [\Phi_{n+1}^{\rm particle}(\vec{r}_{N+1},\cdots, \vec{r}_M)]^*
\ee
We note that the hole in $\Phi_n$ produces an excitation with a positive charge of $1/(2n+1)$ whereas the hole  in $\Phi_{n+1}$ produces an excitation with a negative charge of $-1/(2n+1)$ under reverse vortex attachment\cite{Jain07}, as required for consistency under p-h transformation. Similar arguments apply to the particle excitation of composite fermions. (Note that all charges quoted here are with respect to the uniform ground state.) For states with many particle hole excitations of composite fermions, the spatial locations of the particles (holes) of $\Phi_n$ are correlated with the spatial locations of the particles (holes) of the $[\Phi_{n+1}]^*$.  However, for states with several excitations, the situation becomes more complicated, because in general several states exist with the same quantum numbers, which produce several eigenstates for a given interaction. Within the CF theory, accurate wave functions for these eigenstates can be obtained by the method of CF diagonalization \cite{Mandal02}. The above rule for p-h transformation relates an eigenstate to its hole partner.

\subsection{P-h transformation for the CF Fermi sea: Son's conjecture}

Next we apply these ideas to the limit $n\rightarrow \infty$ and show how the situation greatly simplifies. In this limit,  Eq.~\ref{phtrans} reduces to
\be
\Theta\,\Phi_{\rm FS}(\vec{r}_1,\cdots, \vec{r}_N) \,\Theta^{-1}= \Phi_{\rm FS}^*(\vec{r}_{1},\cdots, \vec{r}_N)
\ee
where we have used the same coordinates on both sides because the number of coordinates is the same. (Remember that the coordinates are all dummy variables in Eq.~\ref{fockwf}.) Furthermore, following Rezayi and Read \cite{Rezayi94}, we write the Fermi sea wave function as 
\be
\Phi_{\rm FS}(\vec{r}_1,\cdots, \vec{r}_N)={\rm Det}\left[ e^{-i\vec{k}_i\cdot \vec{r}_j}\right]
\label{Slater}
\ee
where $\{\vec{k}_i\}$ are the occupied states, as appropriate for the torus geometry. 
The p-h transformation now implies:
\be
\Theta\,{\rm Det}\left[ e^{-i\vec{k}_i\cdot \vec{r}_j}\right] \,\Theta^{-1}= {\rm Det}\left[ e^{+i\vec{k}_i\cdot \vec{r}_j}\right]
\label{Son}
\ee
This provides a microscopic derivation for Son's conjecture:

\vspace{3mm}

\noindent{\em Under p-h transformation, the state of composite fermions with occupation $\{\vec{k}_j \}$ transforms into a state with occupation $\{-\vec{k}_j \}$ at the same energy.}

\vspace{3mm}

\noindent We expect this statement to hold for the ground state of composite fermions, the state with a single particle or hole excitation, and for configurations $\{\vec{k}_j \}$ of composite fermions corresponding to very low energies.  It is not valid for highly excited states of composite fermions. 

\subsection{Absence of backscattering}

In an impressive DMRG calculation, Geraedts {\em et al.} \cite{Geraedts15} demonstrated that $2k_{\rm F}$ backscattering of a single composite fermion is suppressed. To understand this in our approach we use another result by Geraedts {\em et al.} \cite{Geraedts15} that applying $\Theta$ twice to any electronic state gives
$(-1)^{M(M-1)/2}$ times the same state.  
For $\nu=1/2$, we have $M=2N$ and we can write
\be
[\Theta]^2 \,|\nu=1/2\rangle = (-1)^{N} |\nu=1/2\rangle
\label{LevinSon}
\ee
Consider now a CF Fermi sea with an odd $N$ such that a single composite fermion lies in a state with momentum $\vec{K}$ just outside a Fermi sea composed of $N-1$ composite fermions. We will assume that the total momentum is $\vec{K}$, and denote the many-particle state as $|\vec{K}\rangle$ (suppressing the momenta of the composite fermions forming the CF Fermi sea). The above results can be written as $\Theta |\vec{K}\rangle=|-\vec{K}\rangle$ and $\Theta^2 |\vec{K}\rangle=-|\vec{K}\rangle$, which imply, following the standard arguments for Kramers theorem (see for example Ref.~\cite{Bernevig13}), that a p-h symmetric perturbation does not couple $|\vec{K}\rangle$ and $|-\vec{K}\rangle$, i.e., does not cause $2k_{\rm F}$ backscattering. (A p-h symmetric perturbation does not couple any two orthogonal states at $\nu=1/2$ related by p-h transformation $\Theta$, but only with the insight of the CF theory can this result be interpreted in terms of the absence  of $2k_{\rm F}$ backscattering of composite fermions.) This argument generalizes to the situation when the composite fermion at $\vec{K}$ is dressed by CF particle-hole pairs excited out of the CF Fermi sea, so long as the state is orthogonal to its particle-hole conjugate state. Even though we have derived this result for odd $N$, we expect, on physical grounds, that backscattering should be effectively suppressed for arbitrary $N$ for any excited composite fermion that is a sharply defined quasiparticle (i.e. close to an eigenstate), which should be the case provided it is sufficiently close to the Fermi energy.

So far, we have worked directly with the strongly correlated many electron state, interpreting the results in terms of composite fermions. In an effective description that assumes free composite fermions, it is natural to make the assumption\cite{Levin16} that each single composite fermion transforms as $\Theta |\vec{k}\rangle=|-\vec{k}\rangle$ and $\Theta^2 |\vec{k}\rangle=-|\vec{k}\rangle$. These equations are consistent with the transformation conditions on the many particle state listed in Eqs. \ref{Son} and \ref{LevinSon}.  As explained by Levin and Son~\cite{Levin16}, this  leads to a $\pi$ Berry phase for an adiabatic loop around the Fermi circle. The $\pi$ phase also implies an absence of backscattering for a disorder that is symmetric under p-h transformation\cite{Levin16}. However, the validity of the free particle assumption is not obvious for composite fermions away from the Fermi level.

\subsection{Remarks}

Wang and Senthil \cite{Wang15} have motivated the $\pi$ phase by arguing that the LLL projection splits a single vortex off of the composite fermion, converting it into a dipole made of two charge $\pm 1/2$ semions oriented perpendicular to $\vec{k}$. When the composite fermion is taken around the CF Fermi circle, the dipole also completes a rotation producing a phase $\pi$.  Another insight into this issue comes from Haldane's 2004 article \cite{Haldane04}, which derived a connection between the intrinsic anomalous Hall effect and the Berry phase for an adiabatic path around the Fermi circle. Applied to the 1/2 CF Fermi sea and interpreting the Hall conductance at $\nu=1/2$ as an anomalous Hall effect for composite fermions, this implies a $\pi$ Berry phase for non-relativistic composite fermions.  Interestingly, the arguments in Refs.~\cite{Wang15,Haldane04} do not rely, at least in a direct fashion, on the p-h symmetry of the CF Fermi sea. It is also worth noting that the starting point for these arguments as well as the one presented above is the Fermi sea of non-relativistic composite fermions, although Wang and Senthil \cite{Wang15} present a scenario for how LLL projection may endow composite fermions with a two component internal structure.

We noted above that the Eq.~\ref{EqAB} is not exact. This implies that the correspondence $\{\vec{k}_j \}\Leftrightarrow\{-\vec{k}_j \}$ for composite fermions under p-h transformation is not exact. One might ask if it is possible, in the spirit of the Landau Fermi liquid theory, to consider very slightly modified composite fermions for which the correspondence $\{\vec{k}_j \}\Leftrightarrow\{-\vec{k}_j \}$ would be exact, together with an exactly p-h symmetric wave function for the 1/2 state. We have not pursued that issue further. 

It is also worth recalling that p-h symmetry is not a necessary requirement for the existence of the CF Fermi sea. Under experimental conditions, the p-h symmetry is only an approximate symmetry of the Hamiltonian at $\nu=1/2$, because of the presence of some -- and sometimes a significant amount of -- LL mixing.  Furthermore, p-h symmetry is not relevant, even in principle, for all incarnations of the CF Fermi sea. Two examples are obvious. When the spin of the composite fermions is included, the standard CF theory predicts a partially spin polarized CF Fermi sea for sufficiently low Zeeman energies, going into a spin unpolarized CF Fermi sea in the limit when the Zeeman energy vanishes\cite{Park98,Murthy07}. For non-fully spin polarized CF Fermi sea at $\nu=1/2$, p-h symmetry is not relevant, as it relates filling factor $\nu$ to $2-\nu$ for spinful electrons. Experimental evidence exists for partially spin polarized CF Fermi sea\cite{Kukushkin99,Kamburov14c}. Experiments\cite{Melinte00,Freytag02} have shown that the temperature dependence of the spin polarization at $\nu=1/2$ is generally consistent with the simple theory that assumes non-interacting non-relativistic composite fermions in zero effective magnetic field. The second example is that of the Fermi sea at $\nu=1/4$ of composite fermions carrying four vortices, which has been confirmed experimentally\cite{Willett93a,Kamburov14}. P-h symmetry is not relevant at $\nu=1/4$. The arguments relying on p-h symmetry are not relevant either to the partially spin polarized CF Fermi sea at $\nu=1/2$ or to the fully or partially spin polarized CF Fermi sea at $\nu=1/4$, but it would be natural to expect, based on general physical grounds, that the physics of all of these CF Fermi seas ought to be the same.  In particular, Haldane's reasoning \cite{Haldane04}predicts $\pi/2$ phase for each spin component for the spin singlet CF Fermi sea (to the extent that the spin-up and spin-down composite fermions can be treated as independent), and the same phase for the fully spin polarized $1/4$ CF Fermi sea. 

Finally, our microscopic treatment clarifies that the identification of p-h transformation of electrons with the $\{\vec{k}_j\}$$\rightarrow$$\{-\vec{k}_j\}$ transformation of composite fermions is special to $\nu=1/2$, and is not valid for general filling factors $\nu\neq 1/2$.

\section{Dispersion of the ``bare" composite fermion}
\label{sec4}

In the simplest theory, Dirac fermions have a linear dispersion whereas non-relativistic fermions have a quadratic dispersion. The dispersion is not necessarily connected to the nature of composite fermions, but one can still ask what is the dispersion of the underlying composite fermions. As noted above, the dispersion of a single composite fermion can only be deduced by asking if the spectra of interacting electrons are described in terms of Dirac or non-relativistic composite fermions. The different dispersions reflect through the different functional forms for the Landau level energies of the particles: the energy of the $n$th LL of a single Dirac particle is given, in appropriate units, by $\sqrt{n}$, whereas for the non-relativistic particle it is equal to $(n+1/2)$. This, in turn, results in different band structures in the excitation spectrum of the multi-particle state. In particular, the excitation spectrum of non-interacting Dirac fermions consists of many more bands than that of non-interacting non-relativistic fermions.  Consider, for example, the 3/7 state. In the standard CF theory, this maps into $\nu^*=3$ of non-relativistic composite fermions, where the $n=0, 1, 2$ $\Lambda$Ls of composite fermions are fully occupied.  The first excited band is produced from the excitations 2$\rightarrow$3, with energy $\hbar\omega_c^*$ (the CF cyclotron energy). The second excited band at $2\hbar\omega_c^*$ consists of excitations 1$\rightarrow$3, 2$\rightarrow$4 and (2$\rightarrow$3)$^2$, where the superscript denotes the number of excitations. Similarly we get bands at $3\hbar\omega_c^*$, and so on. In the mapping into Dirac composite fermions, the 3/7 FQHE state maps into $\nu^*=3+1/2$, with all Dirac Landau levels with $n\leq 3$ occupied. For the lowest excitation 3$\rightarrow$4, the corresponding band is identical, insofar as counting of states and their quantum numbers are concerned, to the first excited band for the non-relativistic fermions. The structure of higher bands is different, however. For example, excitations 2$\rightarrow$4, 3$\rightarrow$5 and (3$\rightarrow$4)$^2$ all produce {\em distinct} bands for Dirac fermions.

\begin{figure}
\begin{center}
\includegraphics[width=0.45\textwidth]{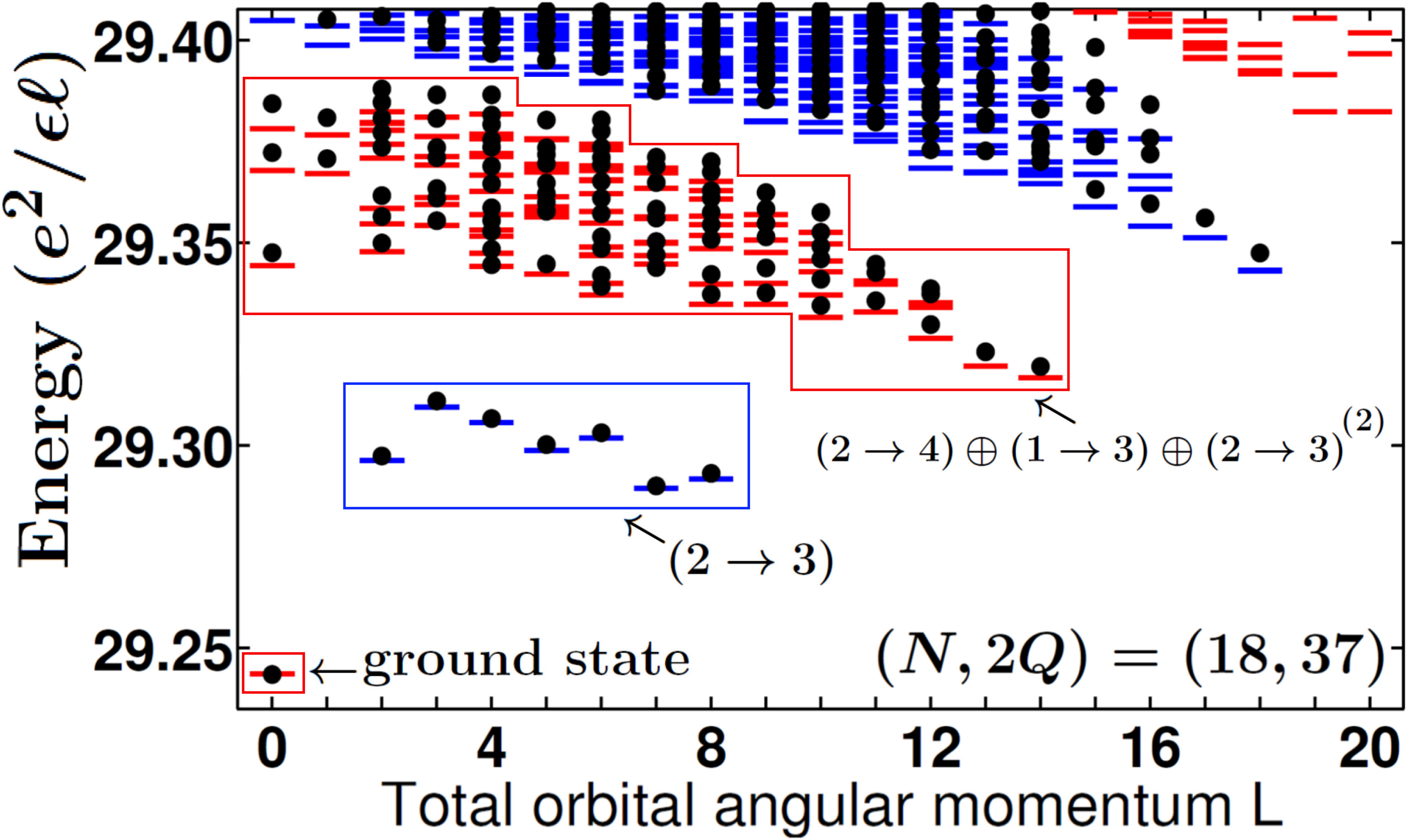}
\caption{(color online) The exact Coulomb spectrum (dashes) of interacting electrons at $\nu=3/7$ for 18 electrons, reproduced from Ref. \onlinecite{Balram13}. Several bands are evident, marked alternately by blue and red colors for ease of depiction. The second excited band (colored blue) is well explained in terms of a combination of the excitations $2\rightarrow 4$, $1\rightarrow 3$ and $(2\rightarrow 3)^2$ of non-relativistic composite fermions. The CF theory based on non-relativistic composite fermions predicts the correct number of states in the band at each quantum number and, furthermore, predicts energies shown by black dots (obtained without any adjustable parameter).}
\label{fig1}
\end{center}
\end{figure}

It is therefore necessary to consider the second or higher excited band to distinguish between Dirac and non-relativistic fermions. Fortunately, the largest system at 3/7 for which exact diagonalization is possible shows well defined first and second excited bands above the ground state, as seen in Fig.~\ref{fig1}. The emergence of the bands itself is highly non-trivial and a result of the formation of composite fermions and their $\Lambda$Ls.  Furthermore, as described in full detail in Ref.~\cite{Balram13}, the description in terms of non-relativistic composite fermions explains the second excited band as a combination of excitations 1$\rightarrow$3, 2$\rightarrow$4 and (2$\rightarrow$3)$^2$. The dots in Fig.~\ref{fig1} show the spectrum predicted from the non-relativistic CF theory.  (It should be noted that the number of states in the actual band is slightly less than that predicted by a mapping into the IQHE of non-interacting particles. Ref.~\cite{Balram13} shows that this discrepancy can be removed by imposing a hard-core constraint on certain tightly bound CF excitons, demonstrating that the discrepancy is thermodynamically insignificant.)  

No splitting into Dirac-CF bands is visible in the second excited band band of Fig.~\ref{fig1}. One may argue, however, that this physics is obscured by the broadening of the bands due to the residual interaction between composite fermions. We therefore ask if it is possible to determine the dispersion of the single composite fermion. This can in principle be accomplished from exact diagonalization studies if one could identify many well defined bands in the excitation spectrum (although this task will be complicated by the fact that the bands have finite widths due to the residual interaction between composite fermions). Such a program is difficult to implement in practice, because of the empirical observation that the bands are well defined approximately up to the energy equal to the Fermi energy of composite fermions, which implies that for the state at $n/(2n+1)$ well defined bands occur for excitations up to $n\hbar\omega_c^*$. Not many bands are therefore visible in the available numerical spectra. One may consider the $n$ dependence of the excitation gaps of different FQHE states along the sequences $n/(2n\pm 1)$, but their interpretation is somewhat complicated by the possible $n$ dependence of the ``CF mass."

\begin{figure}
\begin{center}
\includegraphics[width=0.25\textwidth]{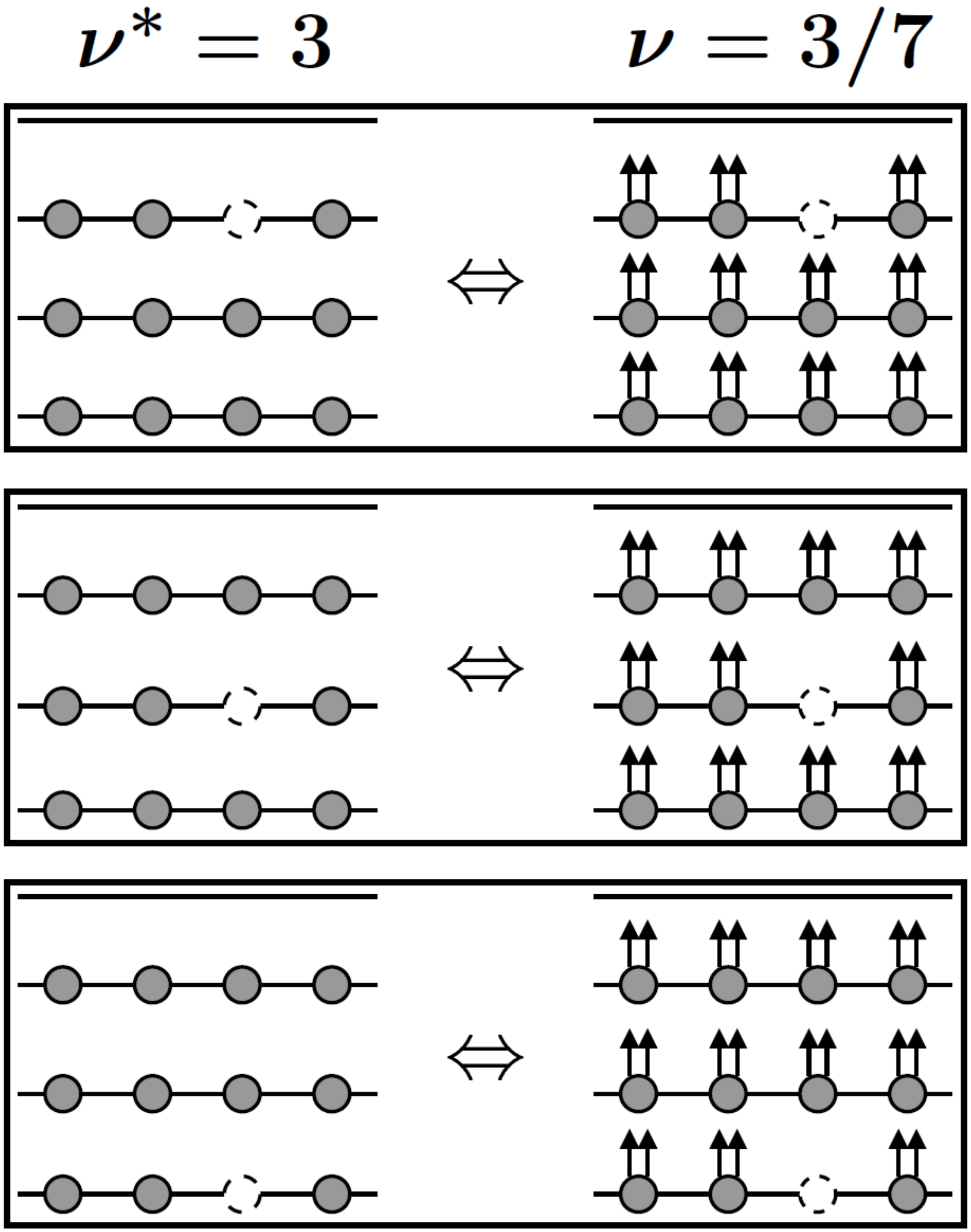}
\includegraphics[width=0.45\textwidth]{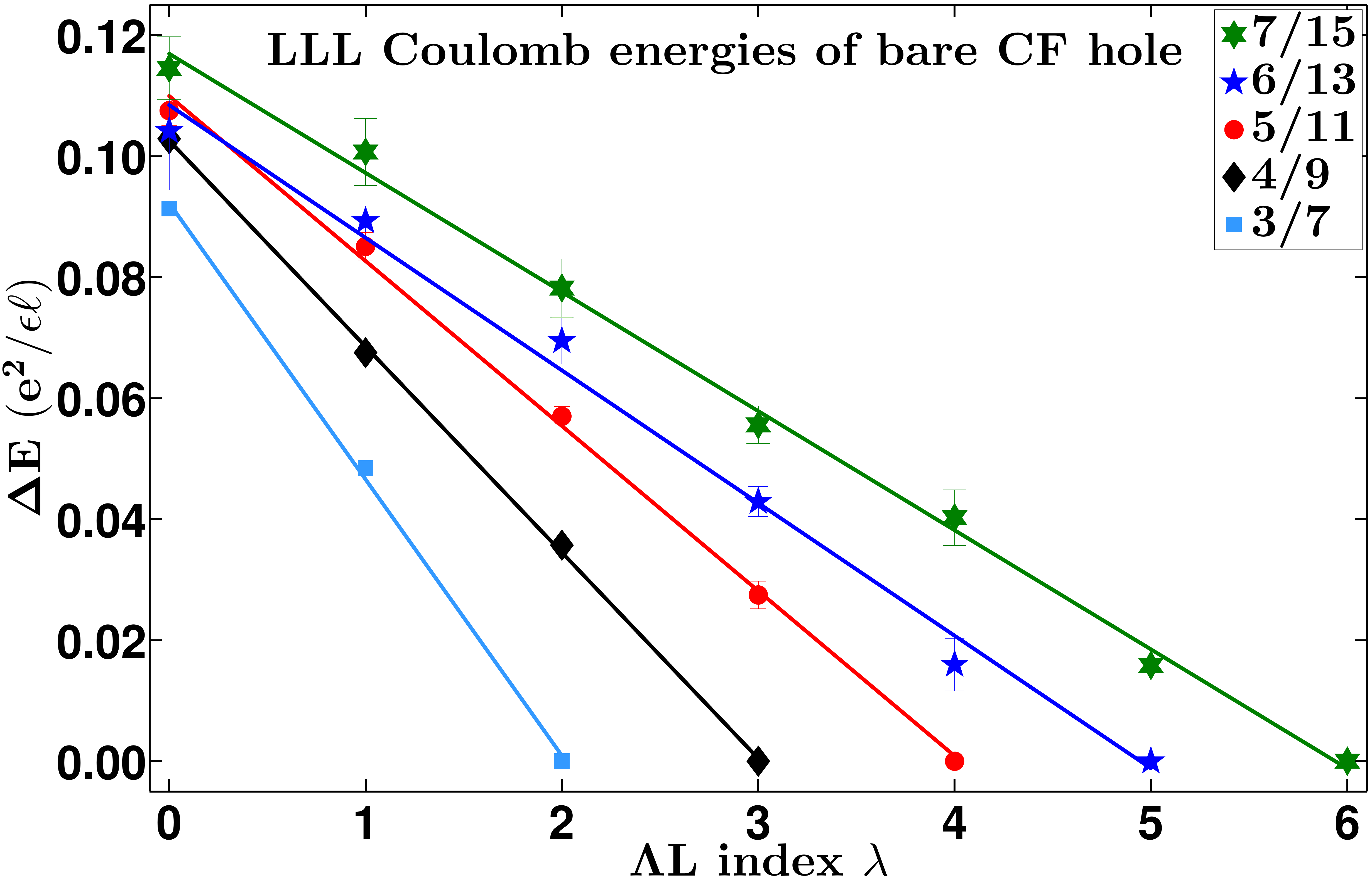}
\caption{(color online) The upper panel shows bare CF hole in various $\Lambda$ levels at $\nu=3/7$ (right panels), analogous to the hole in various Landau levels at $\nu=3$. The lower panel shows the thermodynamic Coulomb energy of the bare CF hole excitation as a function of its $\Lambda$L index for various filling factors in the sequence $n/(2n+1)$. All energies are quoted in Coulomb units of $e^{2}/\epsilon\ell$ and measured relative to the energy of the bare CF hole in the topmost filled $\Lambda$L. The filled symbols are obtained using Jain's wave functions for the CF hole.  The solid straight lines indicate the prediction if composite fermions are taken as non-relativistic fermions.}
\label{fig:CFH_excitation_energy}
\end{center}
\end{figure}

We therefore take the following route. We calculate the energy of the ``bare" CF hole as a function of its $\Lambda$L index $\lambda$. The bare CF hole is obtained by removing a single composite fermion from a filled $\Lambda$L, as shown schematically in Fig.~\ref{fig:CFH_excitation_energy}. 
Fig. \ref{fig:CFH_excitation_energy} shows results for the CF hole energies in different $\Lambda$Ls for up to $\nu=7/15$, where each point represents the thermodynamic limit for the energy of the bare CF hole. This extends a previous calculation by Mandal and Jain\cite{Mandal01}. 
It is evident that the actual bare CF hole energies obtained from the microscopic calculation are consistent with a constant $\Lambda$L spacing, and thus with a parabolic dispersion for the underlying composite fermion. We note that we are using wave functions that are based on a mapping to non-relativistic composite fermions, but we consider this to be a valid procedure given that these wave functions have been shown quantitatively to describe the actual ground and excited states with extremely good accuracy. Furthermore, use of these CF hole wave functions does not guarantee by any means that the spacing between the $\Lambda$Ls would be constant.

We note that the y-intercept in the lower panel of Fig.~\ref{fig:CFH_excitation_energy}, which is the difference between the energy of a hole in the lowest and the highest occupied $\Lambda$Ls, is equal to the CF Fermi energy for sufficiently large $n$.  Indeed, even for $n\geq 4$, this difference is more or less independent of $n$. Our calculations show that the CF Fermi energy is approximately equal to $0.1~e^2/\epsilon\ell$, consistent with the number reported in Ref. \cite{Mandal01}. 

An equivalent way of calculating the dispersion would be to evaluate the energy of a CF particle placed in higher unoccupied $\Lambda$Ls. This is more challenging for technical reasons. Empirically, the $\Lambda$Ls are well defined in the exact spectra only for excitation energies less than the Fermi energy, which means that for the $n/(2n+1)$ state we can only consider approximately $n$ $\Lambda$Ls above the Fermi energy. As a result, one must consider large values of $n$ for a definitive conclusion. Unfortunately, our projection method becomes unstable beyond 8 or 9 $\Lambda$Ls because of the large degree of the derivatives, thus making the CF particle excitations not very useful for the question at hand.

Is it possible that the CF dispersion may become linear for FQHE states very close to 1/2? While we cannot rule it out, we do not see any reason to suspect that would be the case. We do not see any such tendency as the filling factor is changed from 3/7 to 7/15.

We end this section with an important caveat: We have only considered here the {\em bare} CF hole. It is not close to an eigenstate of the LLL Coulomb Hamiltonian (except when it lies at the Fermi energy), and will be dressed by CF excitons to acquire a finite quasiparticle width that increases with its energy. Unfortunately, the fully dressed CF hole is a very complicated object, a proper discussion of which is beyond the scope of the present work. Nonetheless, if the energy of the bare CF hole matches with the peak of the quasiparticle spectral weight, the results presented above are meaningful.

\section{LLL projection and p-h symmetry in Chern-Simons theory}
\label{sec5}

The CS theory of Lopez and Fradkin\cite{Lopez91} and HLR \cite{Halperin93} implements the physics of non-relativistic composite fermions by performing a singular gauge transformation that binds point flux quanta to each electron.  The composite fermions in this approach are topologically similar to the composite fermions defined previously in the context of the microscopic theory, but not identical.  The CS mean-field (CS-MF) wave function is given by:
\begin{equation}
 \Psi_{\nu=\frac{n}{2pn\pm 1}}^{\rm CS-MF}=\prod_{j<k} \Bigg ( \frac{z_{j}-z_{k}}{|z_{j}-z_{k}|}\Bigg)^{2p}\Phi_{\pm n}(B^{*}),
\label{CS_wf} 
\end{equation}
where $\Phi_{\pm n}(B^{*})$ is evaluated at the effective magnetic field and contains the Gaussian factor $\exp[-\sum_j |z_j|^2/\ell^{*2}]$ with $\ell^*=\sqrt{\hbar c/e|B^*|}$. The wave function can also be written as $ \Psi_{\nu=\frac{n}{2pn\pm 1}}^{\rm CS-MF}=(\Phi_1/|\Phi_1|)^{2p}\Phi_{\pm n}(B^{*})$, where each factor $\Phi_\nu$ contains the Gaussian factor $\exp[-\sum_j|z_j|^2/4\ell_\nu^2]$, where $\ell_\nu=\sqrt{\hbar c/e|B_\nu|}=\sqrt{|\nu|/2\pi \rho}$ with $B_\nu=\rho\phi_0/\nu$. We will follow this convention below. This form turns out to be convenient for generalization to Haldane's spherical geometry\cite{Haldane83}, which we use for all our calculations below. Certain elementary facts about the spherical geometry are given in Appendix \ref{sphere_geo}.

The CS-MF wave function is unsatisfactory for a number of reasons\cite{Jain07}. First, it has the same probability amplitude as the non-interacting IQHE state at $\nu^{*}=\pm n$ and hence does not build any repulsive correlations between the electrons (besides what is mandated by the Pauli principle). Second, it has a significant admixture with higher LLs as evidenced by the fact that it contains large powers of $\bar{z}$'s through the factors $\Phi_{\pm n}(B^{*})$ and $|z_{i}-z_{j}|^{2p}$. The approach in Ref.~\onlinecite{Jain89} was to make modifications in the CS-MF wave function to eliminate these deficiencies while retaining its topological character (i.e. the winding phases), which led to the wave functions in Eq.~\ref{Jainwf}.

\begin{table*}
\centering
\begin{tabular}{|c|c|c|c|c|c|}
\hline
\multicolumn{1}{|c|}{$\nu$} & \multicolumn{1}{|c|}{N}   & \multicolumn{1}{|c|}{dim$_{L_{z}=0}$} & \multicolumn{1}{|c|}{dim$_{L=0}$} & \multicolumn{1}{|c|}{$|\langle \Psi^{\rm CF}|\Psi^{\rm Coulomb} \rangle|^{2}$} & \multicolumn{1}{|c|}{$|\langle {\cal P}_{\rm LLL}  \Psi^{\rm CS-MF}|\Psi^{\rm Coulomb} \rangle|^{2}$} \\ \hline
1/3		&	4	&	18	&	2	&	0.99608	&	0.9858(0)  \\ \hline
		&	5	&	73	&	2	&	0.99812	&	0.9879(0)  \\ \hline
		&	6	&	338	&	6	&	0.99289	&	0.9747(0)  \\ \hline
		&	7	&	1656	&	10	&	0.99273	&	0.9697(1)  \\ \hline
		&	8	&	8512	&	31	&	0.99082	&	0.9481(7)  \\ \hline
2/5		&	6	&	58	&	3	&	0.99960	&	0.9994(0)  \\ \hline
		&	8	&	910	&	8	&	0.99920	&	0.9969(0)  \\ \hline
2/3		&	6	&	18	&	2	&	0.99307	&	0.9961(0)  \\ \hline
		&	8	&	73	&	2	&	0.99640	&	0.9981(0)  \\ \hline
		&	10	&	338	&	6	&	0.98810	&	0.9928(0)  \\ \hline
$B^*=0$		&	9	&	910	&	8	&	0.99876	&	0.9963(1)  \\ \hline
\end{tabular}
\caption {Overlaps between the exact Coulomb ground states and the corresponding LLL projected Chern-Simons mean field wave functions ${\cal P}_{\rm LLL}  \Psi^{\rm CS-MF}$ are given in the last column, where $\Psi^{\rm CS-MF}$ is defined in Eq.~\ref{CS_wf}. For comparison, the overlaps of the exact Coulomb ground states with the corresponding wave functions $\Psi^{\rm CF}$ of Eq.~\ref{Jainwf} are reproduced from Fano {\em et al.} \cite{Fano86} ($\nu=1/3$), Dev and Jain \cite{Dev92a} ($\nu=2/5$ and $B^*=0$) and Wu {\em et al.}\cite{Wu93} ($\nu=2/3$). The spherical geometry is used. $N$ is the number of particles. The uncertainty from Monte Carlo sampling is shown in the parentheses. Also shown for reference are the total Hilbert space dimension (dim$_{L_{z}=0}$) and the zero total orbital angular momentum subspace dimension (dim$_{L=0}$).}
\label{tab:Table_overlaps} 
\end{table*}

We now ask if the CS field theory is compatible with LLL constraint.  In order for the perturbation theory to produce a LLL wave function, it must effectively renormalize the polynomial part of the uncorrelated wave function $\Phi_n(B^*)$ of the gauge transformed particles into a strongly correlated form $\prod_{j<k}[|z_j-z_k|/(z_j-z_k)]^{2p} \Psi(\{z_j\}) \exp[+\sum_j |z_j|^2/\ell^{*2}]$, in order to produce the LLL wave function $\Psi(\{z_j\})$ containing the correct Gaussian factor for electrons. Within the field theoretical method itself, it has not been possible to identify the Feynman diagrams that will produce meaningful LLL results. We therefore ask a more modest question: Can the CS-MF wave function be projected into the LLL and, if so, how well does the projected wave function compare to the exact solution? The CS-MF wave function does not have the standard form that we expect from a valid wave function in a magnetic field, i.e., it is not a polynomial of $z_j$'s and $\bar{z}_j$'s multiplying the Gaussian factor. It is thus not immediately obvious how, or even if, the wave function can be projected into the LLL.

We therefore seek to project the CS-MF wave function into the LLL by brute force for finite systems. 
We proceed by formally writing down the CS-MF wave of Eq. \ref{CS_wf} as:
\begin{equation}
 \Psi_{\nu=\frac{n}{2pn\pm 1}}^{\rm CS-MF}=\sum_{i} c_{i}|i\rangle + \text{extra terms},
\end{equation}
where $|i\rangle$ are orthonormalized many-body basis states (Slater determinants) confined to the LLL, and the extra terms are all orthogonal to every $|i\rangle$. 
The (unnormalized) projected CS-MF wave function is defined as:
\begin{equation}
{\cal P}_{\rm LLL} \Psi_{\nu=\frac{n}{2pn\pm 1}}^{\rm CS-MF}=\sum_{i} c_{i}|i\rangle.
 \label{CSp_wf}
\end{equation}
To explicitly carry out this projection we again use the spherical geometry and determine the expansion coefficients $c_{i}=\langle i|\Psi_{\nu=\frac{n}{2pn\pm 1}}^{\rm CS-MF}\rangle$ 
by performing the multi-dimensional integrals for all LLL basis functions $|i\rangle$ by the Monte Carlo method. Since the number of LLL basis states grows exponentially with the number of particles $N$, we have only been able to perform the projection of the CS-MF state for up to $N=10$. We note that it may be possible to carry out projection for larger system sizes using the energy projection method of Ref. \cite{Fremling16}, but we have not pursued that here. Once we have the explicit state, its overlap with the exact Coulomb ground state can be evaluated straightforwardly. 

Table \ref{tab:Table_overlaps} gives the overlaps of the projected CS-MF state with the exact Coulomb ground state. The overlaps are quite high. We also consider the quasihole (qh) and the quasiparticle (qp) along the  sequence $n/(2n\pm 1)$, following the same procedure as above. As shown in Table \ref{tab:Table_overlaps_CS_es}, the overlaps of the projected CS-MF qp and qh with the exact Coulomb qp and qh  are also reasonably high. (For systems smaller than the ones listed in these tables there is only a single state in the relevant total orbital angular momentum sector and hence the overlaps for these systems are trivially equal to unity.) 

We find the degree of agreement with the exact Coulomb solutions to be quite surprising in view of the fact, stated above, that the CS-MF wave functions do not build any repulsive correlations between particles. It appears that provided one begins with wave functions with the correct topological phase factors, the act of LLL projection itself induces strong correlations between particles, attesting to the robustness of the CF physics. In particular, these results suggest that if a LLL projection could be implemented in the CS theory for the $\nu=1/2$ Fermi sea, it would produce a state that honors p-h symmetry to a good approximation.

\begin{table*}
\centering
\begin{tabular}{|c|c|c|c|c|}
\hline
\multicolumn{1}{|c|}{$\nu$} & \multicolumn{1}{|c|}{N}   & \multicolumn{1}{|c|}{dim$_{L_{z}=L}$} & \multicolumn{1}{|c|}{dim$_{L=L^{\rm qp/qh}}$}  & \multicolumn{1}{|c|}{$|\langle {\cal P}_{\rm LLL}  \Psi^{\rm CS-MF}|\Psi^{\rm Coulomb} \rangle|^{2}$} \\ \hline
1/3 qp	&4	&11	&2		&0.9690(0)  \\ \hline
	&5	&46	&3		&0.9723(0)  \\ \hline
	&6	&217	&12		&0.9458(0)  \\ \hline
1/3 qh	&4	&23	&3		&0.9690(0)  \\ \hline
	&5	&98	&5		&0.9723(0)  \\ \hline
	&6	&464	&18		&0.9459(0)  \\ \hline
	&7	&2306	&53		&0.9406(1)  \\ \hline
2/3 qp	&7	&23	&3		&0.9862(0)  \\ \hline
	&9	&98	&5		&0.9918(0)  \\ \hline
	&11	&464	&18		&0.9775(1)  \\ \hline
	&13	&2306	&53		&0.9571(8)  \\ \hline
2/3 qh	&5	&11	&2		&0.9856(0)  \\ \hline
	&6	&46	&3		&0.9961(0)  \\ \hline
	&7	&217	&12		&0.9763(1)  \\ \hline
	&11	&1069	&31		&0.9746(3)  \\ \hline
2/5 qh	&7	&282	&10		&0.9844(0)  \\ \hline
	&9	&4890	&76		&0.9673(1)  \\ \hline
\end{tabular}
\caption {Overlaps of the projected Chern-Simons-mean-field wave functions ${\cal P}_{\rm LLL}  \Psi^{\rm CS-MF}$ -- with the corresponding exact Coulomb ground states for quasihole (qh) and quasiparticle (qp) excitations at 1/3, 2/3 and 2/5 in the spherical geometry. Also shown for reference are the total Hilbert space dimension (dim$_{L_{z}=L}$) and the $L=L^{\rm qp/qh}$ subspace dimension (dim$_{L=L^{\rm qp/qh}}$), where $L^{\rm qp/qh}$ is the total orbital angular momentum quantum number at which the qp/qh occurs.}
\label{tab:Table_overlaps_CS_es} 
\end{table*}

\section{LLL projection for generalized wave functions}
\label{LLLgen}

In the wave functions of Eq.~\ref{Jainwf}, prior to the LLL projection, the number of vortices attached to the each electron, as counted through the winding phases, also is equal to the number of ``zeroes" attached to it (not counting the Pauli zero). The CS wave function of Eq.~\ref{CS_wf}, on the other hand, attaches to each electron $2p$ vortices but no zeroes.  A generalized wave function was introduced in Ref.~\cite{Kamilla97b} which we reproduce here (in the spherical geometry): 
\begin{equation}
 \chi_{\frac{n}{2pn\pm 1}}(\alpha)
 =\frac{\Phi_{1}^{2p}}{|\Phi_{1}|^{2p\alpha}}\Phi_{\pm n}
 \label{gen-wf}
\end{equation}
where the number of attached vortices is $2p$ and the number of attached zeroes is $2p(1-\alpha)$. This wave function reduces to the unprojected Jain wave function for $\alpha=0$ and to the CS-MF wave function for $\alpha=1$, but is defined for other values of $\alpha$ as well (except for $\alpha>(2p+1)/2p$, when the wave function becomes ill-defined: here $\chi(\alpha)$ diverges when two electrons approach one another). The change in the vortex structure does not affect the monopole strength at which these wave functions occur. Because $|\Phi_{1}|^{2p\alpha}=[\Phi_{1}^{*}]^{p\alpha}\Phi_{1}^{p\alpha}$ and the monopole strength of $[\Phi_{1}^{*}]$ is precisely negative of that of $\Phi_{1}$, the denominator makes no contribution to the monopole strength. The above wave function occurs at a monopole strength of $2Q=4pQ_{1}\pm 2Q_{n}=2p(N-1)\pm (N-n^2)/n$ with $2Q_{n}=(N-n^2)/n$, independent of $\alpha$. 

The overlaps of the projected versions of these generalized wave functions with the exact LLL Coulomb ground states are shown in Table \ref{tab:Table_overlaps_gen_CF} for certain values of $\alpha$.  These are again quite high, further confirming the view that provided we begin with a wave function with the correct topology, the LLL projection itself produces good correlations. 

\begin{table*}[htpb]
\centering
\begin{tabular}{|c|c|c|c|c|c|c|}
\hline
\multicolumn{1}{|c|}{$\nu$} & \multicolumn{1}{|c|}{N}   & \multicolumn{1}{|c|}{dim$_{L_{z}=0}$} & \multicolumn{1}{|c|}{dim$_{L=0}$} & \multicolumn{1}{|c|}{$|\langle{\cal P}_{\rm LLL} \chi(1/2)|\Psi^{\rm C} \rangle|^{2}$} & \multicolumn{1}{|c|}{$|\langle {\cal P}_{\rm LLL}\chi(-1/2)|\Psi^{\rm C} \rangle|^{2}$} & \multicolumn{1}{|c|}{$|\langle {\cal P}_{\rm LLL}\chi(-1)|\Psi^{\rm C} \rangle|^{2}$} \\ \hline
1/3	&	4	&	18	&	2	&	0.9919(0)	&	0.9985(1)	&	0.9996(0)	\\ \hline
	&	5	&	73	&	2	&	0.9946(0)	&	0.9989(1)	&	0.9975(4)	\\ \hline
	&	6	&	338	&	6	&	0.9858(0)	&	0.9918(10)	&	0.9900(17)	\\ \hline
	&	7	&	1656	&	10	&	0.9825(5)	&	0.9854(12)	&	0.9743(50)	\\ \hline
	&	8	&	8512	&	31	&	0.9706(8)	&	0.9498(51)	&	0.9562(41)	\\ \hline
2/5	&	6	&	58	&	3	&	0.9995(0)	&	0.9989(2)	&	0.9969(5)	\\ \hline
	&	8	&	910	&	8	&	0.9967(14)	&	0.9864(22)	&	0.9704(89)	\\ \hline
2/3	&	6	&	18	&	2	&	0.9947(0)	&	0.9902(6)	&	0.9849(20)	\\ \hline
	&	8	&	73	&	2	&	0.9973(0)	&	0.9926(4)	&	0.9821(22)	\\ \hline
	&	10	&	338	&	6	&	0.9903(1)	&	0.9645(53)	&	0.9454(134)	\\ \hline	
$B^*=0$	&	9	&	910	&	8	&	0.9967(9)	&	0.9828(31)	&	0.9708(32)	\\ \hline
\end{tabular}
\caption {Overlaps of the LLL projected $\chi(\alpha)$, defined in Eq. \ref{gen-wf}, with the exact Coulomb ground states $\Psi^{\rm C}$ in the spherical geometry. The uncertainty from Monte Carlo sampling is shown in the parentheses. Also shown for reference are the total Hilbert space dimension (dim$_{L_{z}=0}$) and and the zero total orbital angular momentum subspace dimension (dim$_{L=0}$).}
\label{tab:Table_overlaps_gen_CF} 
\end{table*}

\section{Quasiparticle charge and adiabatic continuity}
\label{sec6}

One may ask if the unprojected CS-MF state in Eq.~\ref{CS_wf} is perturbatively connected to the LLL projected state in Eq.~\ref{CFFS}. An important measure of the topological structure of this state is the charge of its quasiparticle or quasihole. We can evaluate this charge for the CS-MF state in two ways. In both derivations, we assume, as usual, that the ground state, quasiparticle and quasihole at $\nu=n/(2pn+1)$ are related to the ground state, quasiparticle and quasihole at $\nu^*=n$ through Eq.~\ref{CS_wf}. 

First, we note that the ground state, quasiparticle and quasihole have exactly the same density profiles as the corresponding states at $\nu^*=n$. It therefore follows that the local charge, i.e. the charge deficiency or excess associated with the quasihole or quasiparticle, is precisely equal to the electron charge $e$ in magnitude. This can also be understood from the fact that the addition of a composite fermion requires adding an electron and two flux quanta. Because the flux in the CS-MF approach carries no charge (i.e. creates no correlation hole), the charge of the electron flux bound state is simply equal to the electron charge.

Alternatively, we can determine the charge of a single quasiparticle by determining how many quasiparticles are created when a single electron is added to the system. (Removal of an electron to produce quasiholes can be considered analogously.) This is most easily addressed in the spherical geometry. As shown in Ref.~\cite{Jain07}, the answer to this calculation depends only on the relation between $2Q$ and $2Q^*$, which is $Q=Q^*+p(N-1)$ independent of $\alpha$. From this, one can deduce, following the discussion in Ref.~\cite{Jain07}, that $2pn+1$ quasiparticles are created when an electron is added at a fixed $Q$, which can be taken to imply that the charge of each is of magnitude $e/(2pn+1)$. 

\begin{figure}
\begin{center}
\includegraphics[width=0.45\textwidth]{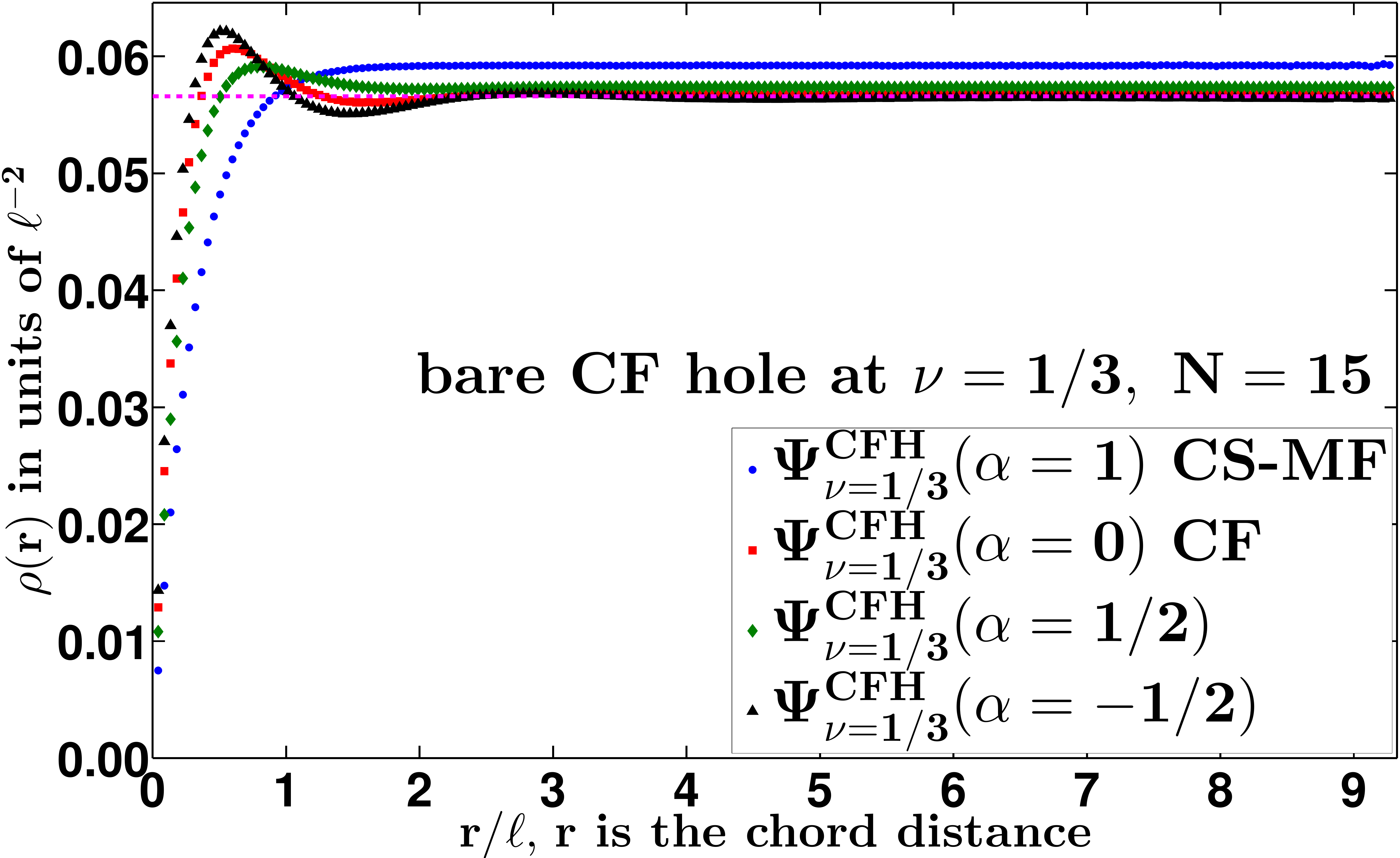}
\caption{(color online) Comparison of the non-normalized density of the generalized CF quasihole state at $\nu=1/3$ for different values of $\alpha$ for $N=15$ electrons. The coordinate $r$ is the chord distance on the sphere. The horizontal dotted magenta line shows the density of the corresponding uniform incompressible state.}
\label{fig:density_compare_diff_alpha}
\end{center}
\end{figure}

How do we resolve this apparent contradiction? The error in the above argument is in the last sentence of the preceding paragraph, which implicitly assumes that the charge of the background state is not altered when we add an electron. In reality, when an electron is added, $2pn+1$ quasiparticles indeed are created, but each has a unit charge $-e$, with charge $2pne$ spread uniformly in the background state. The situation for holes is analogous. This is seen in Fig. \ref{fig:density_compare_diff_alpha}, where we show the density profiles of the quasihole state at $\nu=1/3$ for different values of $\alpha$, defined in Eq.~\ref{gen-wf}. For $\alpha=0$ (red curve) the charge density away from the quasihole goes to the uniform background charge density (dotted magenta line), so the local charge of the quasihole is equal to a third of the electron charge. Contrast this with the $\alpha=1$ Chern-Simons quasihole (blue dotted curve) where the charge deficiency associated with the quasihole is one electron charge, but the density far from the quasihole is slightly higher than the uniform background charge density, which precisely accounts for the remaining $-2/3$ charge.

The fact that the quasiparticle charges of the CS-MF state and the projected CF state are different implies that  the act of LLL projection represents a non-perturbative effect. However, as shown in Ref.\cite{Lopez91}, random phase approximation (RPA) in the CS theory produces the correct quasiparticle charge, suggesting that the CS theory combined with RPA can be adiabatically connected to the projected solution.

We define the natural value for the parameter $\alpha$ in Eq.~\ref{gen-wf} by requiring that the wave function $\prod_{j<k}[(z_j-z_k)^{2p}/ |z_j-z_k|^{2p\alpha}]  \Phi_{\pm n}(B)$ produce the correct filling factor $\nu=n/(2pn\pm 1)$, where the factor $\Phi_{\pm n}(B)$ is evaluated at the external magnetic field (which appears in the Gaussian factor) and $\Phi_{-n}(B)= [\Phi_n(B)]^*$. Noting that this wave function has the same density as $\prod_{j<k}(z_j-z_k)^{2p(1-\alpha)}  \Phi_{n}(B)$, we calculate its filling factor by dividing $N$ by the largest power of, say, $z_1$, which is $2p(1-\alpha)N+N/n$ and equate it to $n/(2pn\pm 1)$ to determine $\alpha$. This yields the following ``normal" form:
\begin{equation}
\chi^{\rm unproj-normal}_{\nu= {n\over 2pn\pm 1}}=
\begin{cases}
\prod_{j<k}(z_j-z_k)^{2p}\Phi_{n}(B) , & \nu={n\over 2pn+1}\\
\prod_{j<k}
{(z_j-z_k)^{2p} \over |z_j-z_k|^{2/n}} 
[\Phi_{n}(B)]^*, & \nu={n\over 2pn-1}\\
\end{cases}
\label{normal}
\end{equation}
[In the spherical geometry, the normal form is given by $\Phi_{1}^{2p}\Phi_{n}$ for $ \nu={n\over 2pn+1}$, and ${\Phi_{1}^{2p} |\Phi_{1}|^{-2/n}} (\Phi_{n})^*$ for $\nu={n\over 2pn-1}$.]
This normal form for the unprojected wave functions has many nice properties. First of all, it explicitly has the correct Gaussian factor and the correct filling factor. Further, it produces the correct quasiparticle and quasihole charge even for the unprojected wave functions. Let us take two examples. For $\alpha=0$, the unprojected 2/3 state $\prod_{j<k}(z_j-z_k)^2 \Phi_2^*$ has the same density as the 2/5 state, and thus has filling factor 2/5 and quasiparticle charge 1/5. Although this unphysical feature is corrected upon projection,  the unprojected wave function $\prod_{j<k}{(z_j-z_k)^{2} \over |z_j-z_k|}  [\Phi_{2}]^*$ already has the correct charge and correct filling factor even without projection. Another interesting example is $\prod_{j<k}(z_j-z_k)^2 \Phi_1^*$ which has the filling factor 1/3 before projection, but 1 after projection, in contrast to $\prod_{j<k}{(z_j-z_k)^{2} \over |z_j-z_k|^2}  [\Phi_1]^*$ which has filling factor 1 both before and after projection. Finally, the normal form is very nicely consistent with the theoretical exponent that describes the power law decay of the edge Green function as well as the theoretical tunneling exponent, which is $\alpha=2p+1$ for $\nu=n/(2pn+1)$ and $\alpha=2p+1-2/n$ for the fractions $\nu=n/(2pn-1)$~\cite{Wen90,Kane94,Shytov98}.

To summarize, the value of $\alpha$, which determines the structure of the quantized vortices, is not particularly important if the wave function is projected into the LLL. For this purpose, the wave functions with $\alpha=0$ are the most convenient because they can be projected  with the Jain-Kamilla projection method for fairly large systems\cite{Jain97,Jain97b,Moller05,Davenport12}. For the unprojected wave functions, on the other hand, the normal form in Eq.~\ref{normal} gives the correct answers for the topological charge of the quasiparticles and quasiholes.

\section{Concluding remarks}
\label{sec7}

Son's proposal that composite fermions are Dirac particles was motivated by certain deficiencies of the Chern-Simons formulation of composite fermions, especially the lack of p-h symmetry of the $\nu=1/2$ state. We note that many of these deficiencies are not shared by the microscopic theory of non-relativistic composite fermions. In particular, the microscopic theory produces an essentially particle-hole symmetric wave function for the CF Fermi sea.  We ask which features of the Dirac-CF approach can be rationalized within the microscopic theory of non-relativistic composite fermions.

We demonstrate that, within the microscopic theory of non-relativistic composite fermions, the p-h transformation for electrons at $\nu=1/2$ translates into a time-reversal-like $\{\vec{k}_j\}$$\rightarrow$$\{-\vec{k}_j\}$ transformation on composite fermions, as assumed by Son \cite{Son15}.  Supplemented with the identity $\Theta^2=(-1)^N$ valid at $\nu=1/2$ \cite{Geraedts15}, this implies absence of $2k_{\rm F}$  backscattering, for a p-h symmetric perturbation, of a single composite fermion quasiparticle for odd $N$, and presumably for all low energy excited composite fermions for arbitrary $N$. If one further assumes that each composite fermion individually satisfies $\Theta^2=-1$, a $\pi$ Berry phase is produced for an adiabatic loop around the CF Fermi circle \cite{Levin16}. 

We ascertain the dispersion of the underlying composite fermion by considering a bare (undressed) CF hole in different $\Lambda$Ls. We find that the $\Lambda$L separation is independent of the $\Lambda$L index to a good approximation, which is consistent with a parabolic dispersion for the composite fermion. 

We also address the question of whether the CS theory is amenable to LLL projection. We find that a brute force projection of the CF-mean field wave function produces surprisingly accurate wave functions for the FQHE states as well as the Fermi sea. These results suggest (but do not prove) that the lack of particle-hole symmetry may not be an intrinsic difficulty with the Chern-Simons formulation but a technical problem stemming from the lack of the lowest Landau level projection. 

\section{Acknowledgments} We thank M. Barkeshli, S. Kivelson, R. Mong, E. Rezayi, S. Sachdev, R. Shankar, M. Shayegan, V. Shenoy, S. Simon, A. Vishwanath for discussions, and are especially grateful to D. Haldane, C.-X. Liu, G. Murthy, T. Senthil and Dam Son for their valuable insights and comments. We thank S. Das Sarma for encouraging us to write up this work and for comments on the manuscript. This work was supported by the U. S. National Science Foundation Grant no. DMR-1401636. We acknowledge the Research Computing and Cyberinfrastructure at Pennsylvania State University which is in part funded by the National Science Foundation Grant No. OCI-0821527. Some of the numerical calculations were performed using the DiagHam package, for which we are grateful to its authors.

\appendix

\section{Haldane's spherical geometry}
\label{sphere_geo}

For completeness, we list some elementary facts of Haldane's spherical geometry\cite{Haldane83} used for our calculations. This geometry consider $N$ electrons on the surface of a sphere subjected to a radial magnetic flux of $2Q\phi_0$ (where $2Q$ is a positive integer and $\phi_0=hc/e$ is quantum of flux), arising from a monopole located at the center of the sphere. The quantity $Q$ is called the monopole strength, which can take either integer or half integer values. The $n$th LL is the shell with single particle orbital angular momentum $l=Q+n$ and has a degeneracy of $2l+1$. These $2l+1$ degenerate states are labeled by the $z$-component of the single particle orbital angular momentum $m=-l,-l+1,\cdots,l$. The $N$-particle state with $n$ filled LLs occurs at a flux of $2Q=(N-n^2)/n$ and has a total orbital angular momentum $L$ and its $z$-component $L_{z}$ both equal to zero.

The wave functions of Eq.~\ref{Jainwf} take the form $\Phi_\nu={\cal P}_{\rm LLL}\Phi_1^2\Phi_{\nu^*}$ in the shperical geometry. Here $\Phi_1=\prod_{j<k}(u_{j}v_{k}-u_{k}v_{j})$, where $u=\cos(\theta/2)e^{i\phi/2}$ and $v=\sin(\theta/2)e^{-i\phi/2}$ are the spinor coordinates, $\theta$ and $\phi$ are the polar and azimuthal angles on the sphere respectively, and $\mathcal{P}_{\rm LLL}$ implements the LLL projection. If the effective monopole strength at $\nu^*$ is $Q^*$, then the monopole strength at $\nu$ is given by $Q=Q^*+N-1$, which implies the standard relation between $\nu$ and $\nu^*$.

\section{Wave functions in the disk geometry}
\label{diskwf}

At first, there appears to be a difference between the disk and the spherical geometries. The generalized wave functions in the disk geometry are written as 
\begin{equation}
 \Psi_{\nu=\frac{n}{2pn\pm 1}}^{\rm un. gen. CF}(\alpha)=\prod_{j<k} \frac{(z_{j}-z_{k})^{2p}}{|z_{j}-z_{k}|^{2p\alpha}}\Phi_{\pm n}(B_{\rm eff}).
\label{eq_un_gen_CF_disk}
\end{equation}
The effective magnetic field is obtained by demanding that the wave function be at the right filling factor. Doing so we find:
 \begin{equation}
 B_{\rm eff}(\alpha)=\frac{B(\beta n\pm 1)}{2pn\pm 1},~~~\beta=2p(1-\alpha) 
 \label{eq_Beff}
 \end{equation}
 where $\beta$ is defined as the number of zeros on an electron in addition to the single zero mandated by the Pauli exclusion principle (i.e. the wave function vanishes as $r^{\beta+1}$). To avoid divergences in the wave function as two particles approach each other we need $\beta+1\geq 0 \Rightarrow \alpha \leq 1+(2p)^{-1}$.  The fact that the effective magnetic field depends on $\alpha$ appears to be distinct from that in the spherical geometry. 
 
However, we can eliminate these differences if we define the wave function in the disk geometry slightly differently:
\begin{equation}
\Psi_{\nu=\frac{n}{2pn\pm 1}}^{\rm un. gen. CF}(\alpha)=\frac{\Phi_1^{2p}}{|\Phi_1|^{2p\alpha}}\Phi_{\pm n}(B_{\rm eff}').
\label{eq_un_gen_CF_sphere}
\end{equation}
where the wave function of one filled LL is given by:
\begin{equation}
\Phi_{1}=\prod_{j<k}(z_{j}-z_{k})e^{-\frac{1}{4\ell_{1}^2}\sum_{i}|z_{i}|^2},~~~\ell_{1}=\sqrt{\frac{\hbar c}{eB\nu}}
\end{equation}
The relation between $B'_{\rm eff}$ and $B$ can be obtained by equating the Gaussian factors of the wave functions given in Eqs. \ref{eq_un_gen_CF_disk} and \ref{eq_un_gen_CF_sphere} and making use of the value of $B_{\rm eff}(\alpha)$ given in Eq. \ref{eq_Beff}. Doing so we get
\begin{equation}
B_{\rm eff}'(\alpha)=\Big( \frac{\pm 1}{2pn\pm 1} \Big)B=B_{\rm eff}(\alpha=1)
\end{equation}
which is independent of $\alpha$ and the same as $B^{*}$, the effective magnetic field seen by composite fermions.

\bibliography{../../Latex-Revtex-etc./biblio_fqhe}
\bibliographystyle{apsrev}
\end{document}